\documentclass[a4paper,11pt]{article}
\usepackage{jcappub}
\usepackage[T1]{fontenc}
\usepackage{lmodern}
\usepackage[utf8]{inputenc}
\DeclareUnicodeCharacter{2223}{\ensuremath{\mid}}
\usepackage{bm,booktabs,array,siunitx}
\usepackage[section]{placeins}
\graphicspath{{figures/}}

\title{Propagation delays and regional intensity changes in lensed hotspot images}
\author{Yong Li,}
\author{Xiao-Gang Lan}
\affiliation{\href{https://www.cwnu.edu.cn/}{School of Physics and Astronomy, China West Normal University, Nanchong 637009, China}}
\emailAdd{xglan@cwnu.edu.cn}
\abstract{Propagation delays cause an image recorded at a single observer time to combine radiation emitted at different stages of the source evolution. Comparable changes in total intensity can accompany distinct and even opposite changes in apparent image size. Using regional intensities, centroids, and covariances, we apply the law of total covariance to separate changes in regional intensity weights from changes in internal widths and centroid separation. Ray tracing simulations of a finite Gaussian hotspot moving along a prescribed strong field trajectory show two events with comparable attenuation in screen integrated intensity but opposite changes in second moment size. In the contracting event, the regional weights move away from balance; in the expanding event, they move toward balance even as the centroids approach each other. Differential propagation delays therefore drive these opposite size responses by redistributing intensity between spatially separated image regions.}

\hypersetup{
  pdftitle={Propagation delays and regional intensity changes in lensed hotspot images},
  pdfauthor={Yong Li, Xiao-Gang Lan},
  pdfsubject={Propagation delay effects in lensed finite hotspot images},
  pdfkeywords={black hole hotspot; time dependent imaging; propagation delay; regional intensity redistribution; image moments; relativistic ray tracing}
}
\begin{document}
\maketitle
\raggedbottom
\section{Introduction}

A spatially resolved image of an evolving relativistic source combines signals emitted at different times in its history. Photons arriving at the same observer time can originate from different stages of the source motion and carry information about different source states at different screen positions. Such temporal mixing is particularly important when the source varies on a timescale comparable to the spread in photon travel times \cite{RojasPaternina2026,Bronzwaer2018RAPTOR}. Horizon scale images of M87* and Sagittarius A* (Sgr A*) \cite{EHTM87,EHTSgrA} and near infrared astrometry of Galactic center flares \cite{GRAVITY2018} make this temporal mismatch relevant to strong field observations. General relativistic magnetohydrodynamic (GRMHD) simulations associate localized, evolving emission near black holes with magnetic reconnection, magnetic flux eruptions, and nonthermal flaring \cite{Ripperda2020,Chatterjee2021Flares,LinYuan2024Flares}.

By separating source motion, gravitational lensing, and arrival time effects under controlled conditions, hotspot models have been widely used to calculate images, light curves, spectra, and centroid tracks for compact orbiting emitters \cite{CunninghamBardeen1973,SchnittmanBertschinger2004,Broderick2005,BroderickLoeb2006,Vincent2011,Baubock2020,Huang2024,Chen2024,WuChen2024CPC}. The same framework also accommodates noncircular trajectories and their time dependent image morphology \cite{Huang2024}. Reconnection driven plasmoids and pattern motion can likewise produce localized, evolving emission structures \cite{Ripperda2020,Matsumoto2020Pattern}. Recent calculations have quantified propagation time differences between image orders \cite{Kocherlakota2024}, while photon ring studies have examined how magnification and image morphology change with image order \cite{Kocherlakota2024,Gralla2019}. Time dependent radiative transfer calculations relax the fast light approximation by evaluating the source state at the emission time of each segment along a ray \cite{Bronzwaer2018RAPTOR,Dexter2009,Pelle2022Skylight}. A recent comparison of propagation prescriptions interprets the resulting images through the competition between source variability and the distribution of lensing delays \cite{RojasPaternina2026}. Slow light calculations have also revealed substantial changes in the time dependent, resolved morphology of the M87 jet \cite{Tsunetoe2026}.

Our focus is how differential temporal sampling redistributes intensity between spatially separated lensed regions and thereby determines whether the second moment size of a finite hotspot image decreases or increases.

Although a spatially uniform multiplicative attenuation leaves the image centroid and normalized second moment unchanged, a nonuniform response can alter the image size, with the sign of the change determined by where intensity is suppressed or enhanced. Reducing the relative weight of screen regions that contribute strongly to the covariance generally favors contraction, whereas increasing it favors expansion. Changes in regional widths and centroid separation can further reinforce or offset this tendency. A single mean delay cannot distinguish these scale responses because it removes the association between the source history and screen position. To interpret the response in terms of temporal sampling, we retain this association between the delay history and screen position and characterize each separated image region by its intensity weight, centroid, internal covariance, and transfer response. Regional weights are updated through these response factors. The law of total covariance separates the contribution of the weight changes from changes in regional widths and centroid separation.

We use ray traced images of a finite comoving Gaussian hotspot moving along a prescribed strong field trajectory to determine whether regional intensity redistribution dominates the opposite size responses and how it competes with changes in regional widths and centroid separation. Six frozen states trace the evolution of the separated image regions. Two retarded events with comparable total attenuation but opposite size changes provide examples of contraction and expansion. Both are analyzed using image moments. Complete regional transfer histories are available for the contracting event, while a resolution dependence study accompanies the expanding event. Supplementary calculations compare frozen backgrounds at a fixed instantaneous source state and, in a separate set of independent frozen and retarded image pairs, examine the effects of inclination and source size.

Section 2 describes regional intensity redistribution and image size, and Section 3 describes the numerical implementation. Section 4 presents the contraction and expansion results and their regional interpretation, while Sections 5 and 6 discuss the physical implications and summarize the main conclusions. Model details and numerical checks are collected in Appendices A and B.

\section{Regional intensity redistribution and image size}

\subsection{A moving finite source}

All calculations use geometric units \(G = c = M = 1,\) where \(M\) is the black hole mass, with the source rest mass set to unity in the particle Hamiltonian. Lengths and times are reported in units of \(M.\) The background is the static, spherically symmetric Bardeen geometry \cite{Bardeen1968,AyonBeato2000},

\setcounter{equation}{0}
\begin{equation}
ds^{2} = - f(r)dt^{2} + \frac{dr^{2}}{f(r)} + r^{2}(d\theta^{2} + \sin^{2}\theta\, d\phi^{2}),\quad\quad f(r) = 1 - \frac{2r^{2}}{(r^{2} + g_{m}^{2})^{3/2}},
\end{equation}

where \(g_{m} = 0.5\, M\) is the Bardeen magnetic charge parameter. Source center motion in the prescribed azimuthal potential \(A_{\phi} = \mathcal{B}r^{2}\sin^{2}\theta/2\) is governed by the Hamiltonian

\setcounter{equation}{1}
\begin{equation}
H = \frac{1}{2}g^{\mu\nu}(p_{\mu} - qA_{\mu})(p_{\nu} - qA_{\nu}) = - \frac{1}{2}.
\end{equation}

Source charge is parameterized by \(q\), \(\mathcal{B}\) sets the amplitude of the prescribed external magnetic field, and \(L = p_{\phi}\) is the conserved canonical azimuthal momentum. The Bardeen metric is fixed as the background, and the azimuthal potential is treated as an externally maintained test field. The supporting current and its gravitational backreaction are not evolved. The prescribed field and charged particle Hamiltonian determine the source center worldline, along which the position and velocity evolve rapidly and the emissivity profile is prescribed. For the principal calculation, we use \(q\mathcal{B} = 0.036998688\) and \(L = 3.335\), with a reference event on the outward leg at \(r_{0} = 6.5\, M\). After turning at \(8.41320\, M\), the orbit returns inward and approaches a radial barrier at \(4.32805\, M\), producing a sequence of rapidly changing source positions and velocities within the stored interval. Charged particle motion in prescribed magnetic fields has long served as a controlled model for studying strong field relativistic orbital structure \cite{Wald1974,FrolovShoom2010,FrolovShoomTzounis2014,Kolos2015,Tursunov2016,Panis2019,Rayimbaev2023Regular}. Source states are labeled by the unwrapped azimuthal advance

\setcounter{equation}{2}
\begin{equation}
\psi = \phi - \phi_{entry},
\end{equation}

The continuously unwrapped azimuth is denoted by \(\phi,\) with \(\phi_{entry}\) setting the phase origin at the selected inward entry event. Integration begins earlier at \(r_{0} = 6.5\, M\) on the outward leg; Appendix A specifies the entry criterion. This phase convention applies to all six overview images and both principal events. Along the stored worldline, the emissivity pattern remains continuously defined.

A comoving Gaussian hotspot profile \cite{BroderickLoeb2006} with a finite numerical cutoff describes the source. Its source centered form is defined through a local tangent space coordinate constructed from the tetrad projected separation and the local source velocity,

\setcounter{equation}{3}
\begin{equation}
D_{com}^{2} = \delta_{\widehat{\imath}\widehat{\jmath}}\xi^{\widehat{\imath}}\xi^{\widehat{\jmath}} + \gamma^{2}(v_{\widehat{\imath}}\xi^{\widehat{\imath}})^{2},\quad\quad\gamma = (1 - v^{2})^{- 1/2},
\end{equation}

where \(v^{2} = \delta_{\widehat{\imath}\widehat{\jmath}}v^{\widehat{\imath}}v^{\widehat{\jmath}},\) and the velocity dependent term accounts for the component of the separation parallel to the source motion. The comoving emissivity profile is

\setcounter{equation}{4}
\begin{equation}
j_{com} = j_{c}\exp\left( - \frac{D_{com}^{2}}{2\sigma_{s}^{2}} \right),\quad\quad\sigma_{s} = R_{s}/2,\quad\quad D_{com} \leq 3R_{s}.
\end{equation}

The central comoving emissivity and Gaussian width are specified by \(j_{c}\) and \(\sigma_{s} = R_{s}/2,\) respectively. Applying the cutoff \(D_{com} \leq 3R_{s}\) truncates the distant Gaussian tail, with zero emissivity outside this finite support. All principal images use \(R_{s} = M.\) Within each frozen and retarded pair, the comoving width and central emissivity \(j_{c}\) remain fixed; the source evolves through its position and velocity along the prescribed worldline. Source size comparisons use normalized image moments and retarded to frozen intensity ratios, all of which remain invariant under a common rescaling of \(j_{c}\) within each pair.

\subsection{Observer time image formation}

Photon trajectories follow null geodesics in the metric of Eq. (2.1). In the covariant formulation of radiative transfer \cite{Younsi2012}, an optically thin ray segment intersecting the source contributes

\setcounter{equation}{5}
\begin{equation}
dI_{\nu} \propto \mathcal{G}_{\infty}^{3}j_{\nu}\, d\ell_{em},\quad\quad\mathcal{G}_{\infty} = \frac{- k_{t}}{- u_{\mu}k^{\mu}},\quad\quad d\ell_{em} = ( - u_{\mu}k^{\mu})|d\lambda|,
\end{equation}

A positive path length element in the emitter frame is denoted by \(d\ell_{em}\). Gravitational and Doppler frequency shifts enter through \(\mathcal{G}_{\infty}\), with \(u^{\mu}\) denoting the local emitter four velocity and \(k^{\mu}\) the future directed photon momentum. The numerical screen lies at finite radius. The factor \(\mathcal{G}_{\infty}\) gives the redshift to the static observer at infinity used for the reported intensity. Defined in the emitter frame, the gray emissivity \(j_{\nu}\) is frequency independent and follows the prescribed spatial profile \(j_{com}\). This invariant form of optically thin transfer also applies to time dependent general relativistic ray tracing \cite{Bronzwaer2018RAPTOR,Pelle2022Skylight}.

While the frozen calculation holds the source state fixed, each ray in the retarded calculation samples the source worldline at

\setcounter{equation}{6}
\begin{equation}
t_{em} = t_{obs} + t_{\gamma}(\lambda),\quad\quad z_{s} = z_{s}(t_{em}),
\end{equation}

Along a backward integrated ray, \(t_{\gamma}(\lambda) \equiv t(\lambda) - t_{obs} \leq 0\) gives the coordinate time offset. Source position and velocity for the transfer calculation are supplied by the stored phase space state \(z_{s}(t)\). Using the short path that passes closest to the source center as a synchronization anchor fixes a common observer time while preserving the variation in emission time across the screen through the path dependent \(t_{\gamma}.\)

The frozen and retarded calculations use the same screen null geodesics in the fixed background while sampling different source positions and velocities. The sampled state changes the overlap with the moving emissivity profile, the redshift factor \(\mathcal{G}_{\infty}\), and the comoving path length \(d\ell_{em}\) in Eq. (2.6). With zero absorption, these quantities set the regional intensity response.

Summing over pixels with centers \(\mathbf{x}_{ij} = \left( x_{ij},y_{ij} \right)\) and areas \(\Delta A_{ij},\) we obtain the screen integrated intensity and image centroid,

\setcounter{equation}{7}
\begin{equation}
F = \sum_{ij}^{}I_{ij}\Delta A_{ij},\quad\quad\mathbf{C} = \frac{1}{F}\sum_{ij}^{}\mathbf{x}_{ij}I_{ij}\Delta A_{ij},
\end{equation}

and the intensity covariance and second moment size are

\setcounter{equation}{8}
\begin{equation}
\Sigma_{I} = \frac{1}{F}\sum_{ij}^{}I_{ij}(\mathbf{x}_{ij} - \mathbf{C})(\mathbf{x}_{ij} - \mathbf{C})^{\mathsf{T}}\Delta A_{ij},\quad\quad\sigma_{eff} = \sqrt{\operatorname{tr}\,\Sigma_{I}}.
\end{equation}

Intensity weighted second moments also provide image size constraints in VLBI reconstruction \cite{Issaoun2019}. A uniform grid has \(\Delta A_{ij} = \Delta A\). These normalized moments require \(F > 0\) and distinguish uniform attenuation from translation and spatial redistribution.

\subsection{Regional response and weight update}

Mutually exclusive regions \(b\) partition the quantitative image domain and together cover all pixels used in the moment sums. Applying the same partition to the frozen and retarded states ensures that each response \(A_{b}\) is evaluated over the same pixels in both states. For \(s \in \{ fr,ret\},\) let \(F^{s}\) denote the corresponding total screen integrated intensity. Regional intensity and fractional weight are defined as

\setcounter{equation}{9}
\begin{equation}
F_{b}^{s} = \sum_{(i,j) \in b}^{}I_{ij}^{s}\Delta A_{ij},\quad\quad p_{b}^{s} = \frac{F_{b}^{s}}{F^{s}},\quad\quad\sum_{b}^{}p_{b}^{s} = 1.
\end{equation}

For each region, the net retarded to frozen response is defined as

\setcounter{equation}{10}
\begin{equation}
A_{b} = \frac{F_{b}^{ret}}{F_{b}^{fr}},\quad\quad F_{b}^{fr} > 0,
\end{equation}

which describes the response of region \(b\) to the kinematic history sampled by its rays. Substituting this response into Eq. (2.10) gives the updated regional weight,

\setcounter{equation}{11}
\begin{equation}
p_{b}^{ret} = \frac{p_{b}^{fr}A_{b}}{\overline{A}},\quad\quad\overline{A} = \sum_{c}^{}p_{c}^{fr}A_{c} = \frac{F^{ret}}{F^{fr}}.
\end{equation}

Normalization is set by the scalar \(\overline{A}\), which gives the net screen integrated intensity response. Updating the weights requires \(F_{b}^{fr} > 0\) in each retained region and \(F^{ret} > 0;\) their updated values depend on the relative regional response factors.

Within each region, the stored transfer increments retain information about the delay distribution. Delay is measured relative to the direct ray synchronization anchor as \(\tau = \left( - t_{\gamma} \right) - \tau_{anchor}\), where \(\tau_{anchor} \equiv - t_{\gamma,anchor}\) is the anchor propagation delay. Thus \(\tau < 0\) denotes a propagation time shorter than that of the anchor path. A source intersecting segment \(n\) in screen pixel \(\left( i_{n},j_{n} \right)\) contributes the retarded state transfer increment \(\Delta I_{n}\), before multiplication by the pixel area. At relative delay \(\tau_{n}\), the screen integrated weight is therefore \(\Delta F_{n} \equiv \Delta I_{n}\Delta A_{i_{n}j_{n}}\). For a region with \(F_{b}^{ret} > 0\), the corresponding kernel and normalized distribution are

\setcounter{equation}{12}
\begin{equation}
K_{b}(\tau) = \sum_{n:\,(i_{n},j_{n}) \in b}^{}\Delta F_{n}\,\delta(\tau - \tau_{n}),\quad\quad P_{b}(\tau) = \frac{K_{b}(\tau)}{\int K_{b}(\tau')d\tau'}.
\end{equation}

Each transfer increment \(\Delta I_{n}\) incorporates the source state, redshift and comoving emissivity sampled along its ray segment; multiplication by \(\Delta A_{i_{n}j_{n}}\) gives the integrated screen contribution. The normalized history \(P_{b}\) records the distribution of source times represented in region \(b.\) Paired frozen and retarded images provide a direct measurement of the response factor \(A_{b},\) while \(P_{b}\) characterizes the regional delay distribution.

\subsection{Regional intensity weights and image size}

For each region \(b\) with \(F_{b}^{s} > 0,\) we define the centroid and covariance as

\setcounter{equation}{13}
\begin{equation}
\mathbf{C}_{b}^{s} = \frac{1}{F_{b}^{s}}\sum_{(i,j) \in b}^{}\mathbf{x}_{ij}I_{ij}^{s}\Delta A_{ij},\quad\quad\Sigma_{b}^{s} = \frac{1}{F_{b}^{s}}\sum_{(i,j) \in b}^{}I_{ij}^{s}(\mathbf{x}_{ij} - \mathbf{C}_{b}^{s})(\mathbf{x}_{ij} - \mathbf{C}_{b}^{s})^{\mathsf{T}}\Delta A_{ij}.
\end{equation}

Applying the law of total covariance \cite{ChampSills2024} to the regional intensity distribution gives

\setcounter{equation}{14}
\begin{equation}
\Sigma_{I}^{s} = \sum_{b}^{}p_{b}^{s}\Sigma_{b}^{s} + \sum_{b}^{}p_{b}^{s}(\mathbf{C}_{b}^{s} - \mathbf{C}^{s})(\mathbf{C}_{b}^{s} - \mathbf{C}^{s})^{\mathsf{T}}.
\end{equation}

Its trace can be decomposed as

\setcounter{equation}{15}
\begin{equation}
T^{s} \equiv \operatorname{tr}\,\Sigma_{I}^{s} = W^{s} + B^{s},\quad W^{s} = \sum_{b}^{}p_{b}^{s}\operatorname{tr}\,\Sigma_{b}^{s},\quad B^{s} = \sum_{b}^{}p_{b}^{s}\left| \mathbf{C}_{b}^{s} - \mathbf{C}^{s} \right|^{2}.
\end{equation}

Within region variances contribute through their weighted sum \(W\), and \(B\) gives the contribution from the separation of the regional centroids. For two separated regions, the between region covariance depends on the centroid separation and the balance of the intensity weights. Let \(p^{s} \equiv p_{R}^{s}\) be the right region weight and define \(\mathbf{d}^{s} = \mathbf{C}_{R}^{s} - \mathbf{C}_{L}^{s}\). Then

\setcounter{equation}{16}
\begin{equation}
\mathbf{C}^{s} = (1 - p^{s})\mathbf{C}_{L}^{s} + p^{s}\mathbf{C}_{R}^{s},\quad\quad B^{s} = p^{s}(1 - p^{s})\left| \mathbf{d}^{s} \right|^{2}.
\end{equation}

For fixed centroid separation \(\left| \mathbf{d} \right|,\) the balance factor \(p(1 - p)\) is maximal at \(p = 1/2,\) when the two regions have equal intensity. As the weights become unequal at fixed regional geometry, the global centroid moves toward the brighter region and the between region contribution decreases. To isolate the weight change at fixed frozen geometry, let \(p \equiv p_{R}^{fr},\) \(p' \equiv p_{R}^{ret}\) and \(\mathbf{d} \equiv \mathbf{d}^{fr}.\) Equation~(2.12) then gives

\setcounter{equation}{17}
\begin{equation}
p' = \frac{pA_{R}}{(1 - p)A_{L} + pA_{R}},\quad\quad B_{w}' = p'(1 - p')\left| \mathbf{d} \right|^{2} = \frac{p(1 - p)A_{L}A_{R}}{\left\lbrack (1 - p)A_{L} + pA_{R} \right\rbrack^{2}}\left| \mathbf{d} \right|^{2}.
\end{equation}

With the frozen geometry held fixed, the resulting change from regional weights alone is \(\Delta B_{w \mid fr} \equiv B_{w}' - B^{fr}.\) Changes in the total trace \(T\) due to changes in regional intensity weights combine the change in \(B\) with the change in the weighted within region covariances. Under the same change in regional intensity weights, the centroid displacement is

\setcounter{equation}{18}
\begin{equation}
\Delta\mathbf{C}_{w} = (p' - p)\mathbf{d}.
\end{equation}

Centroid motion and image scale change thus characterize the same redistribution through moments of different orders. Since \(B^{s} = \beta^{s}D^{s},\) the endpoint change can arise from a change in the regional balance factor \(\beta,\) the squared centroid separation \(D,\) or both. With \(\beta^{s} = p^{s}\left( 1 - p^{s} \right)\) and \(D^{s} = \left| \mathbf{d}^{s} \right|^{2},\) we write this change as

\setcounter{equation}{19}
\begin{equation}
\begin{aligned}
\Delta B = \Delta B_{wt}^{sym} + \Delta B_{geom}^{sym}, \\
\Delta B_{wt}^{sym} = \frac{\beta^{ret} - \beta^{fr}}{2}(D^{ret} + D^{fr}), \\
\Delta B_{geom}^{sym} = \frac{D^{ret} - D^{fr}}{2}(\beta^{ret} + \beta^{fr}).
\end{aligned}
\end{equation}

The first term vanishes if the balance factor is unchanged, including when the regional weights are unchanged; the second vanishes if the centroid separation is unchanged. This identity provides an algebraic decomposition of the endpoint change into weight and separation terms, allowing for dynamical coupling between them. Allowing the regional widths and centroids to vary, define \(\Delta X \equiv X^{ret} - X^{fr}\) for any endpoint quantity \(X\). With \(T^{fr} > 0\), the trace and fractional size responses are

\setcounter{equation}{20}
\begin{equation}
\Delta T = \Delta W + \Delta B,\quad\quad\frac{\sigma_{eff}^{ret} - \sigma_{eff}^{fr}}{\sigma_{eff}^{fr}} = \sqrt{1 + \frac{\Delta T}{T^{fr}}} - 1.
\end{equation}

The sign of the scale response follows from

\setcounter{equation}{21}
\begin{equation}
\left\{ \begin{aligned}
\Delta B < - \Delta W,  & \text{ contraction}, \\
\Delta B = - \Delta W,  & \text{ no scale change}, \\
\Delta B > - \Delta W,  & \text{ expansion}.
\end{aligned} \right.\
\end{equation}

When the regional geometry changes little, a differential response that shifts \(p\) away from \(1/2\) favors contraction, while a shift toward \(1/2\) favors expansion. Equation (2.22) determines whether changes in the internal profiles preserve or reverse this tendency. Regional contributions are evaluated using partitions that keep the image regions separated. The global trace \(T\) is independent of the partition.

\section{Numerical implementation}

Ray tracing uses an extended version of the public Fortran/OpenMP OCTOPUS framework \cite{Hu2025OCTOPUS}, incorporating a finite comoving emitter, observer time worldline queries and regional transfer history diagnostics. Located at \(r_{obs} = 1000\, M\), the orthographic camera has inclination \(i = 85^{\circ}\) and field of view \(\lbrack - 15\, M,15\, M\rbrack^{2}\). Backward null geodesics are integrated using paired fifth order Fehlberg and seven stage sixth order Butcher schemes \cite{Fehlberg1969,Butcher1964}. Integration stops when a ray reaches the outer event horizon \(r_{eh} = 1.78596778\, M\), escapes to \(r = 1500\, M\), reaches \(200,000\) steps, or fails a state or null Hamiltonian check. Appendix B.1 gives the step acceptance and step size prescriptions.

The source worldline is stored on a fixed proper time grid and interpolated linearly at each coordinate emission time using the shortest wrapped azimuthal difference. The phase label \(\psi\) is accumulated from a continuously unwrapped copy. For the principal events, all queries remain within the stored interval containing the recorded retarded samples. A direct primary ray synchronization anchor is selected for the observer time comparison as described in Appendix B.1.

Figure~1 contains six frozen overview images, all calculated at \(N = 1024\). Quantitative calculations at \(\psi = 1.5\) use the archived composite grid with a registered fine window, whereas the \(\psi = 2.5\) event uses uniform grids with \(N = 512\), \(1024\) and \(2048\). Integrated quantities, regional intensities, centroids and covariances are evaluated directly from the unsmoothed native arrays using their native pixel areas. At \(N = 1024\), the overview at \(\psi = 1.5\) is calculated separately for the morphology sequence in Fig.~1. Appendices B.2 and B.3 describe the composite grid coordinates, area accounting, mask construction and control matching. Morphology displays and comparisons based on the Pearson correlation and relative \(L_{2}\) difference use the same zero padded Gaussian convolution, \(\sigma_{scr} = 0.20\, M\). Figures 14 and 15 instead display the native intensity and residual fields without this convolution. Overview and control panels are normalized to their own peaks, whereas direct comparisons use the reference peak specified in the figure caption. Displayed fields of view are cropped around the image regions where appropriate; moment sums use the native computational domains described above.

Pairwise morphology comparisons use the Gaussian convolved fields \({\widetilde{I}}_{1}\) and \({\widetilde{I}}_{2}\). After screen integral normalization, we evaluate the statistics over the active union

\setcounter{equation}{0}
\begin{equation}
\Omega_{act} = \left\{ \max\left( {\widetilde{I}}_{1}/{\widetilde{I}}_{1,peak}, {\widetilde{I}}_{2}/{\widetilde{I}}_{2,peak} \right) > 0.005 \right\}.
\end{equation}

Morphology comparisons use the area weighted Pearson correlation and the reference normalized relative \(L_{2}\) difference defined in Appendix B.3. Unless stated otherwise, intensity panels use a logarithmic color scale with a lower limit of \(10^{-3}\) of the stated display reference peak. In Fig.~3, the signed residual is \(R = \left( {\widetilde{I}}_{ret} - {\widetilde{I}}_{fr} \right)/{\widetilde{I}}_{fr,peak}\). Both events use \(R_{0} = 0.015\,\max_{\text{two events}}|R| \simeq 0.01425\). A gray dotted outline encloses the union of pixels where either convolved field exceeds \(I_{cut} = 10^{-3}\) of that row's frozen peak. Statistical comparisons use the threshold \(0.005\) in Eq. (3.1).

Complete regional transfer histories are available for the \(\psi = 1.5\) calculation; the expanding event is described by regional delay summaries.

\section{Results}

\subsection{A finite hotspot produces separated, evolving image regions}

Figure~1 presents a frozen sequence showing how the separated image regions evolve along the source trajectory. At \(\psi = 0,\) a compact bright region is accompanied by extended, nearly ring shaped emission, whereas at \(\psi = 0.5,\) the compact bright component lies alongside extended arcuate emission. Two separated regions dominate near \(\psi = 1.0\) and \(1.5\) and become broader and farther apart at \(\psi = 2.0\) and \(2.5.\) With the source profile fixed, these changes arise from the source position and velocity along the prescribed trajectory together with the strong field lens map. The coexistence of direct and more strongly deflected components is consistent with earlier hotspot calculations \cite{CunninghamBardeen1973,BroderickLoeb2006,Huang2024,Kocherlakota2024}.

\begin{figure}[htbp]
\centering
\includegraphics[width=\textwidth,height=0.76\textheight,keepaspectratio]{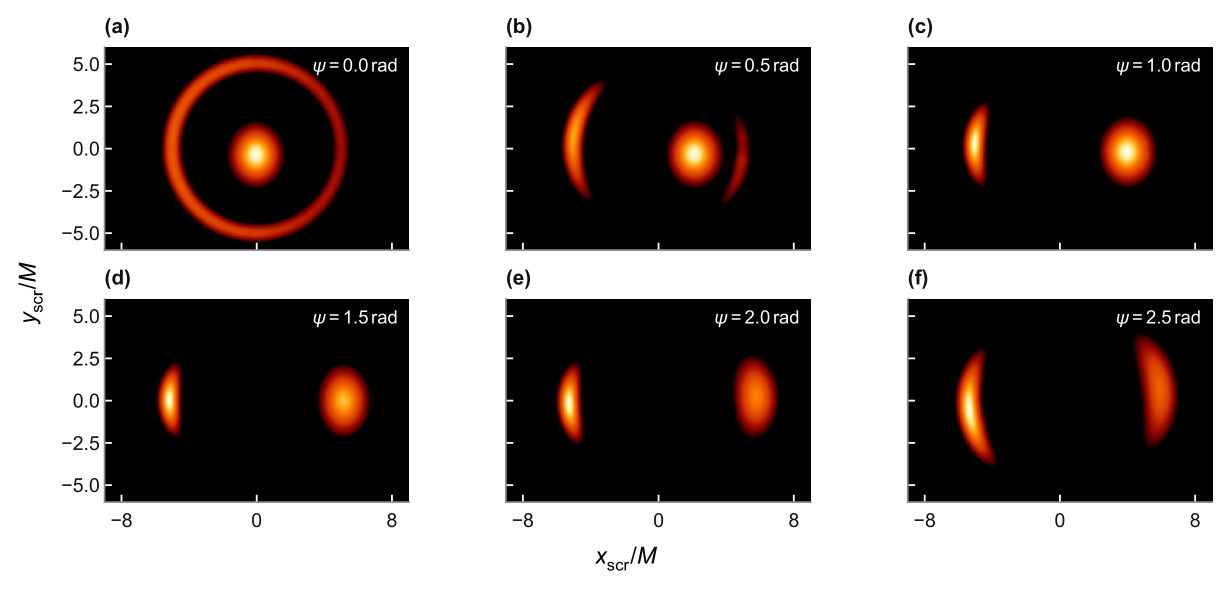}
\caption{Frozen source morphology of the finite comoving Gaussian at six source advances: (a--f) \(\psi = 0,\) \(0.5,\) \(1.0,\) \(1.5,\) \(2.0\) and \(2.5\, rad.\) All panels use the same field of view and logarithmic range, \(10^{-3} \leq I/I_{peak} \leq 1,\) with each normalized to its own peak.}
\label{fig:paper64-1}
\end{figure}

At \(\psi = 1.5\) and \(2.5,\) the between region term dominates the frozen covariance trace. Changes in regional intensity can therefore move the global centroid and alter the apparent size even when the individual regions remain narrow. Table 2 gives the corresponding covariance budgets.

\subsection{Comparable attenuation with opposite scale responses}

Figures 2 and 3 compare the two principal frozen and retarded image pairs. Their screen integrated intensities decline by similar amounts, yet the second moment size decreases by \(44.5\%\) at \(\psi = 1.5\) and increases by \(11.8\%\) at \(\psi = 2.5\) (Table 1). Figure 2 shows the centroid positions for both events; Appendix B.2 quantifies the resolution dependence of the contraction statistics and the fraction of the trace change due to regional weight changes.

\begin{table}[htbp]
\centering
\caption{Principal endpoint responses. Results at \(\psi = 1.5\) use the composite grid, while the expansion row uses \(N = 2048.\)}
\begin{tabular}{lcc}
\toprule
\textbf{Event} & \textbf{Intensity change (\%)} & \textbf{Size change (\%)} \\
\midrule
\(\psi = 1.5\) & \(-57.7\) & \(-44.5\) \\
\(\psi = 2.5\) & \(-61.6\) & \(+11.8\) \\
\bottomrule
\end{tabular}
\end{table}

\begin{figure}[htbp]
\centering
\includegraphics[width=\textwidth,height=0.76\textheight,keepaspectratio]{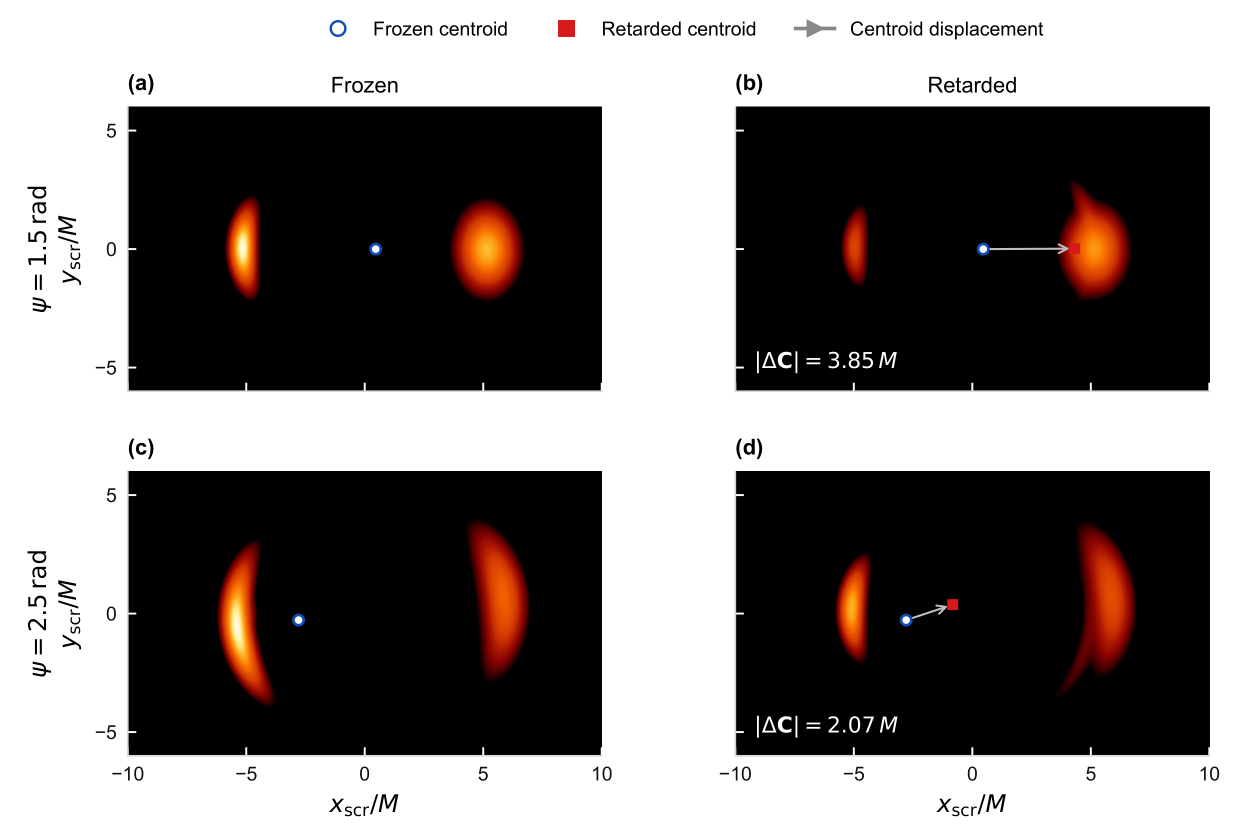}
\caption{Opposite scale responses under comparable dimming. Panels (a, b) show \(\psi = 1.5\, rad\) and panels (c, d) show \(\psi = 2.5\, rad;\) the left and right columns contain the frozen and retarded images, respectively. Both panels in each row use that row's frozen peak and the logarithmic range \(10^{-3} \leq I/I_{fr,peak} \leq 1.\) Blue open circles and red squares mark the frozen and retarded centroids, respectively. Arrows in the retarded panels join the two centroids, while the frozen panels show only the reference position. Figure 3 shows the signed residual contours.}
\label{fig:paper64-2}
\end{figure}

\begin{figure}[htbp]
\centering
\includegraphics[width=\textwidth,height=0.76\textheight,keepaspectratio]{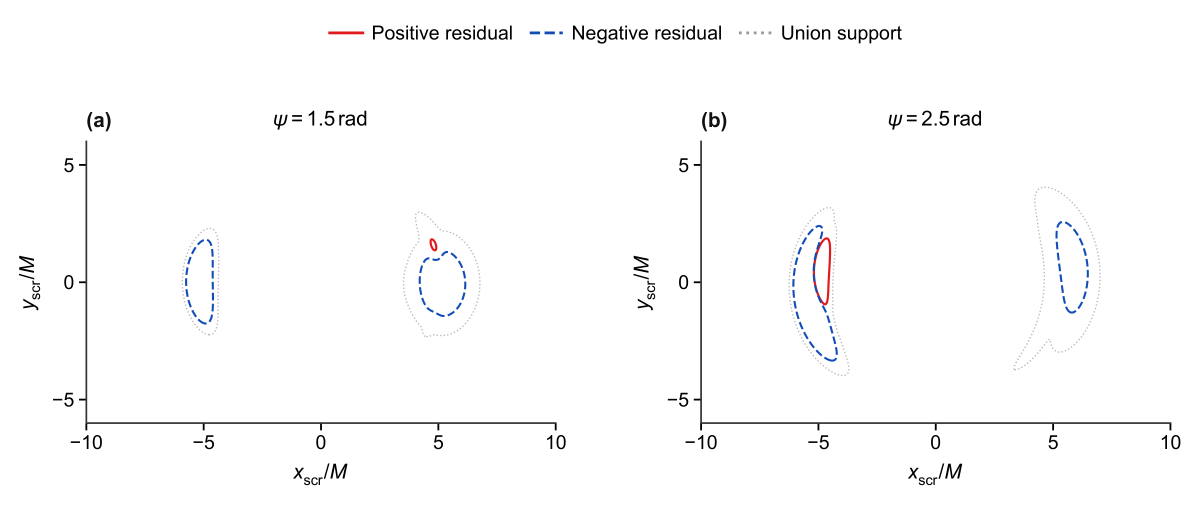}
\caption{Signed residual contours for the high resolution endpoints at (a) \(\psi = 1.5\) and (b) \(\psi = 2.5\, rad.\) For \(R = \left( {\widetilde{I}}_{ret} - {\widetilde{I}}_{fr} \right)/\max{\widetilde{I}}_{fr},\) red solid and blue dashed curves mark \(+ R_{0}\) and \(- R_{0},\) respectively, with \(R_{0} \approx 0.01425\) shared by both events. Gray dotted curves mark the union boundary at \(\max\left( {\widetilde{I}}_{fr}, {\widetilde{I}}_{ret} \right)/\max{\widetilde{I}}_{fr} = 10^{-3}.\) The fields use the same screen convolution, \(0.20\, M,\) as in Fig.~2.}
\label{fig:paper64-3}
\end{figure}

Uniform attenuation leaves both \(\mathbf{C}\) and \(\sigma_{eff}\) unchanged, so the measured changes require spatially nonuniform redistribution. The residual contours and regional intensity fractions show strong suppression of one region in the contracting event, whereas the weaker region gains relative weight in the expanding event.

\subsection{Regional intensity redistribution gives opposite covariance responses}

The right region gains fractional weight in both events, but the initial intensity balances differ (Table 2). At \(\psi = 1.5\), a nearly balanced image becomes dominated by the right region; at \(\psi = 2.5\), the initially weaker right region gains weight and brings the image closer to balance. At fixed regional geometry, these shifts change the between region contribution in opposite directions.

In the contracting event, the decrease in \(B\) exceeds the small increase in \(W,\) giving

\setcounter{equation}{0}
\begin{equation}
\Delta T = \Delta B + \Delta W \approx - 18.449\, M^{2},\quad\quad(\psi = 1.5),
\end{equation}

which satisfies the contraction condition in Eq.~(2.22). In the expanding event, the increase in \(B\) exceeds the decrease in \(W,\) giving

\setcounter{equation}{1}
\begin{equation}
\Delta T = \Delta B + \Delta W \approx 5.729\, M^{2},\quad\quad(\psi = 2.5).
\end{equation}

The second event therefore expands even as its weighted internal term decreases. As a weighted sum, \(W\) depends on both local widths and regional intensity weights.

\begin{table}[htbp]
\centering
\caption{Regional covariance budgets for the contraction and expansion events. The right region weight is denoted by \(p\). Values at \(\psi = 1.5\) use the composite grid, while those at \(\psi = 2.5\) come from the \(N = 2048\) calculation.}
\begin{tabular}{lcccccc}
\toprule
\textbf{Event} & \(p^{fr}\) & \(p^{ret}\) & \(\Delta W/M^{2}\) & \(\Delta B/M^{2}\) & \(\Delta T/M^{2}\) & \textbf{response} \\
\midrule
\(\psi = 1.5\) & 0.543 & 0.918 & \(+ 0.121\) & \(- 18.570\) & \(- 18.449\) & contraction \\
\(\psi = 2.5\) & 0.227 & 0.392 & \(- 0.217\) & \(+ 5.945\) & \(+ 5.729\) & expansion \\
\bottomrule
\end{tabular}
\end{table}

Figure 4 shows Eq. (2.18) as a continuous function of \(A_{R}/A_{L},\) using the frozen weight and centroid separation of each event. With the frozen geometry fixed, the measured ratios set the marked values of \(B_{w}';\) Table~2 reports \(\Delta B\) from the full retarded calculation, which also includes changes in regional geometry.

\begin{figure}[htbp]
\centering
\includegraphics[width=\textwidth,height=0.76\textheight,keepaspectratio]{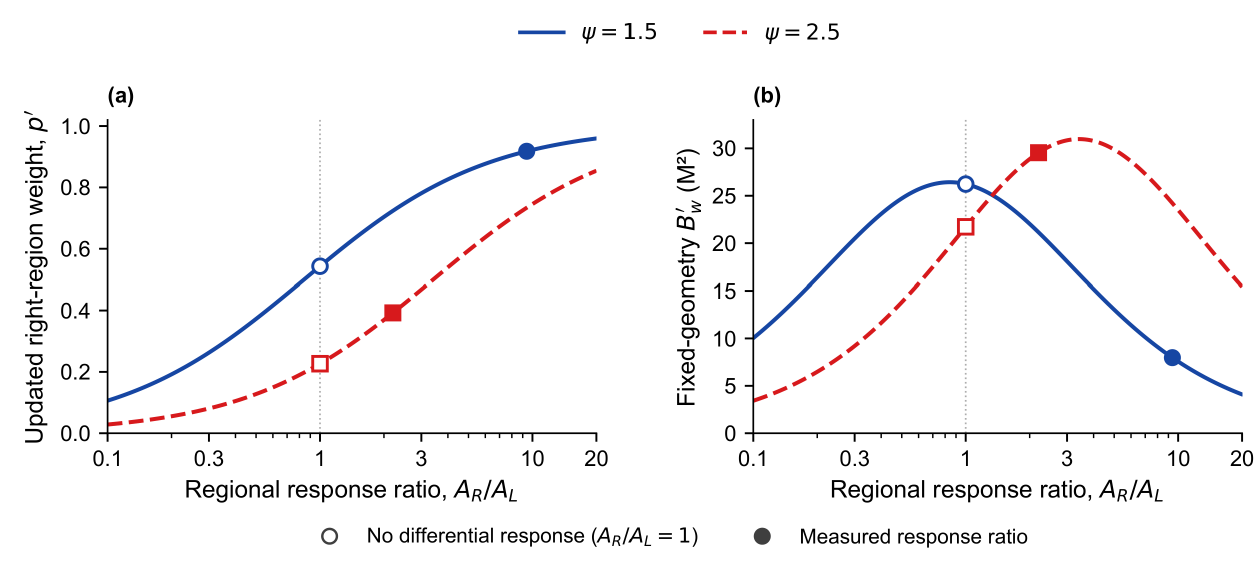}
\caption{Response at fixed frozen geometry from Eq. (2.18). (a) Updated right region weight \(p'\) versus \(A_{R}/A_{L}.\) (b) Corresponding between region contribution \(B_{w}' = p'(1 - p')\left| \mathbf{d}^{fr} \right|^{2}.\) Blue solid and red dashed curves use the frozen weights and separations of the \(\psi = 1.5\) and \(2.5\, rad\) events, respectively. Open markers indicate equal regional responses, \(A_{R}/A_{L} = 1,\) and filled markers indicate the ratios measured from the native image pairs. All markers give analytic values at fixed frozen geometry.}
\label{fig:paper64-4}
\end{figure}

\subsection{Regional intensity weights dominate the major axis response}

Equation (2.20) separates the effects of regional weights and centroid separation. The symmetric weight and geometry contributions to \(\Delta B\) are \(-17.903\) and \(-0.667\,M^{2}\) for \(\psi = 1.5\), and \(+7.572\) and \(-1.627\,M^{2}\) for \(\psi = 2.5\), respectively. In both cases the regional centroids move closer, but at \(\psi = 2.5\) the weight shift more than offsets the geometric contraction. The covariance response is concentrated along the major axis, with much smaller minor axis changes and rotations (Appendix B.5). Changes in regional intensity weights act predominantly along the line joining the separated regions.

\begin{figure}[htbp]
\centering
\includegraphics[width=\textwidth,height=0.76\textheight,keepaspectratio]{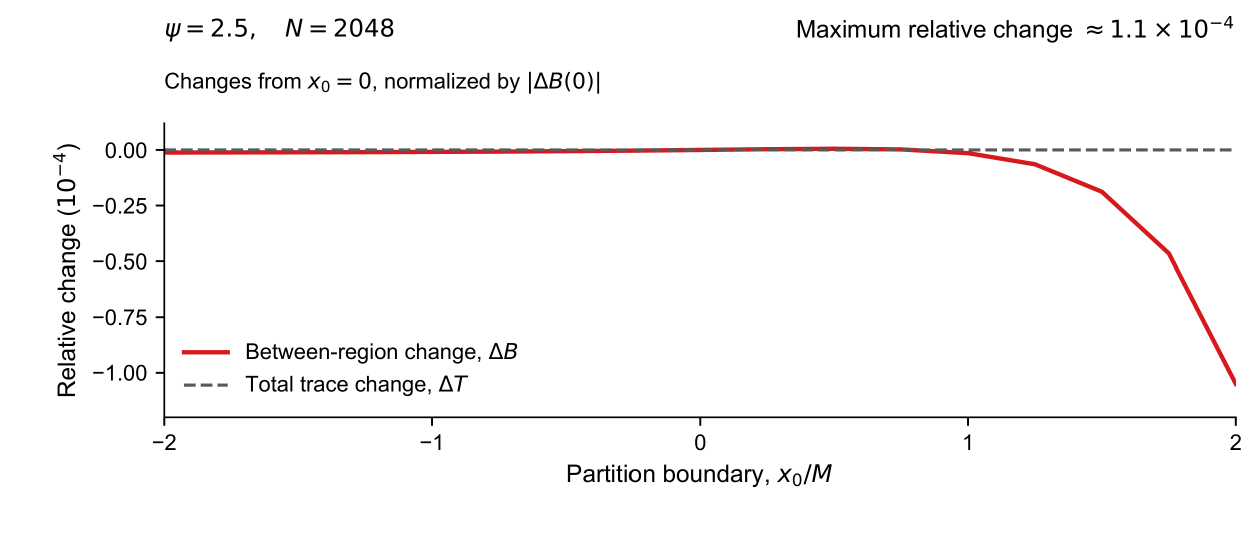}
\caption{Partition sensitivity for the \(\psi = 2.5\, rad,\) \(N = 2048\) expansion pair. The response is evaluated at all 17 recorded boundaries. Red solid and gray dashed curves show the changes in \(\Delta B\) and \(\Delta T\) relative to their values at \(x_{0} = 0,\) each normalized by \(\left| \Delta B(0) \right|.\) The maximum relative change in \(\Delta B\) is about \(1.1 \times 10^{-4},\) while the total trace remains unchanged to numerical roundoff.}
\label{fig:paper64-5}
\end{figure}

Provided the partition still separates the image regions, the expansion budget changes little as the boundary moves across the central gap (Fig.~5). Appendix B.2 gives the contraction mask test.

\subsection{Differential temporal sampling drives contraction at \texorpdfstring{\(\psi = 1.5\)}{psi = 1.5}}

For \(\psi = 1.5\), the stored transfer histories show that the two image regions sample different source times at one observer time. Their weighted median delays differ by \(34.9\,M\), corresponding to \(3.32\,rad\) of unwrapped azimuthal advance along the stored trajectory (Fig.~7(a) and Appendix B.5). The two regions therefore sample substantially different stages of the source motion. The left region is much more strongly suppressed, with \(A_{L}=0.076\) versus \(A_{R}=0.714\), moving the weights away from balance (Table 2). Applied at fixed frozen geometry, these responses give a centroid displacement close to the full retarded value (Appendix B.5).

\begin{figure}[htbp]
\centering
\includegraphics[width=\textwidth,height=0.76\textheight,keepaspectratio]{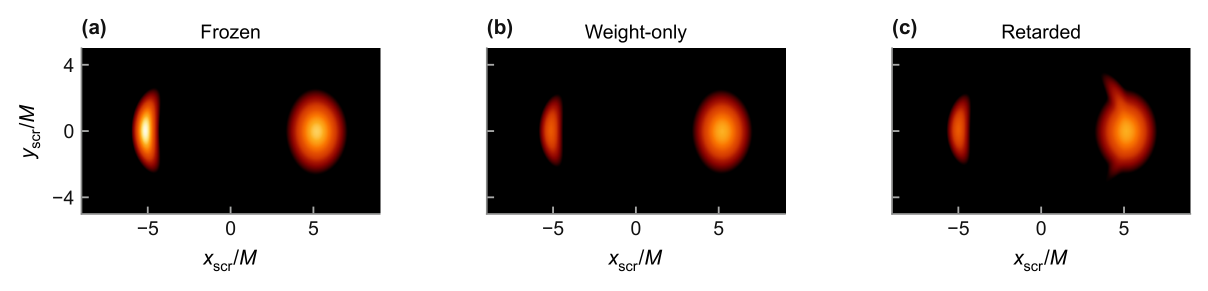}
\caption{High resolution contraction and regional weight reconstruction at \(\psi = 1.5\, rad.\) (a--c) Frozen image, updated regional weights at fixed frozen geometry, and full retarded image. All three images use the frozen peak and logarithmic range \(10^{-4} \leq I/I_{fr,peak} \leq 1.\) Response factors measured from the same composite image pair are used in the reconstruction.}
\label{fig:paper64-6}
\end{figure}

\begin{figure}[htbp]
\centering
\includegraphics[width=\textwidth,height=0.76\textheight,keepaspectratio]{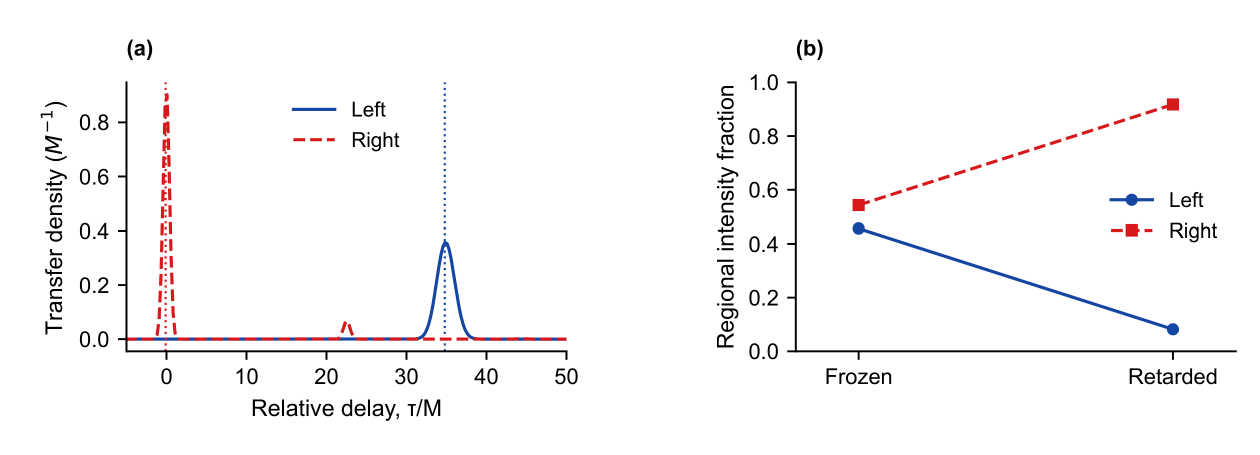}
\caption{Temporal sampling and endpoint weights at \(\psi = 1.5\, rad.\) (a) Retarded state regional transfer weight densities \(P_{b}(\tau),\) sampled every \(0.25\, M\) and plotted as lines. Dotted vertical lines mark the weighted medians, separated by \(34.9\, M.\) (b) Left and right intensity fractions from the composite image pair in Fig.~6. Blue solid lines with circles and red dashed lines with squares denote the left and right regions, respectively.}
\label{fig:paper64-7}
\end{figure}

Applying the two response factors to the frozen regions gives the reconstruction from regional weights alone in Fig.~6(b). Over the active region, its Pearson correlation with the full retarded image is \(0.993\). For this fixed frozen geometry reconstruction, we define the trace fraction \(f_{T}=(T_{w}-T^{fr})/(T^{ret}-T^{fr})\), where \(T_{w}\) is the reconstructed trace. The value \(f_{T}\approx98.6\%\) quantifies the fraction of the total covariance trace reduction accounted for by changes in regional intensity weights. The left and right regional rms sizes change by \(-2.56\%\) and \(+5.64\%\), respectively (Appendix B.5), while the global size decreases by \(44.5\%\). Together with the reconstruction, these values identify the change in relative regional intensity as the main contribution to the contraction.

\begin{figure}[htbp]
\centering
\includegraphics[width=\textwidth,height=0.76\textheight,keepaspectratio]{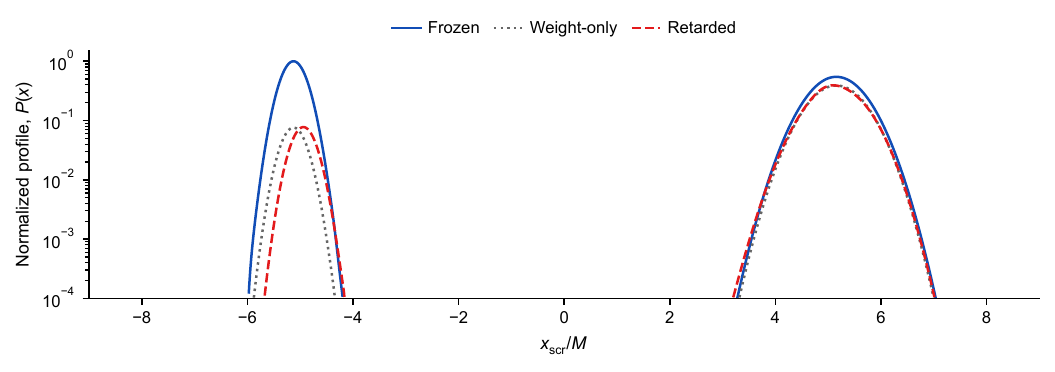}
\caption{Spatial profiles of the frozen, regional weight reconstruction, and full retarded contraction images in Fig.~6. Profiles are defined by \(P(x) = \frac{\int\widetilde{I}(x,y)\, dy}{\max_{x}\int{\widetilde{I}}_{fr}(x,y)\, dy},\) with the same frozen profile denominator for all three curves and a logarithmic vertical axis display limit of \(10^{-4}.\) The Gaussian convolved display fields are used for the profiles; image moments are measured from the native arrays.}
\label{fig:paper64-8}
\end{figure}

\subsection{Differential response shifts the weights toward balance and expands the \texorpdfstring{\(\psi = 2.5\)}{psi = 2.5} image}

At \(\psi = 2.5\), the central dark gap provides a natural partition at \(x=0\). The measured response ratio \(A_{R}/A_{L}=2.20\) increases the right region share toward, but not to, equal weight (Table 2 and Fig.~10(a)). Both regions dim, but the right region dims less. The between region term therefore rises while the weighted internal term falls, producing expansion (Table 2).

\begin{figure}[htbp]
\centering
\includegraphics[width=\textwidth,height=0.76\textheight,keepaspectratio]{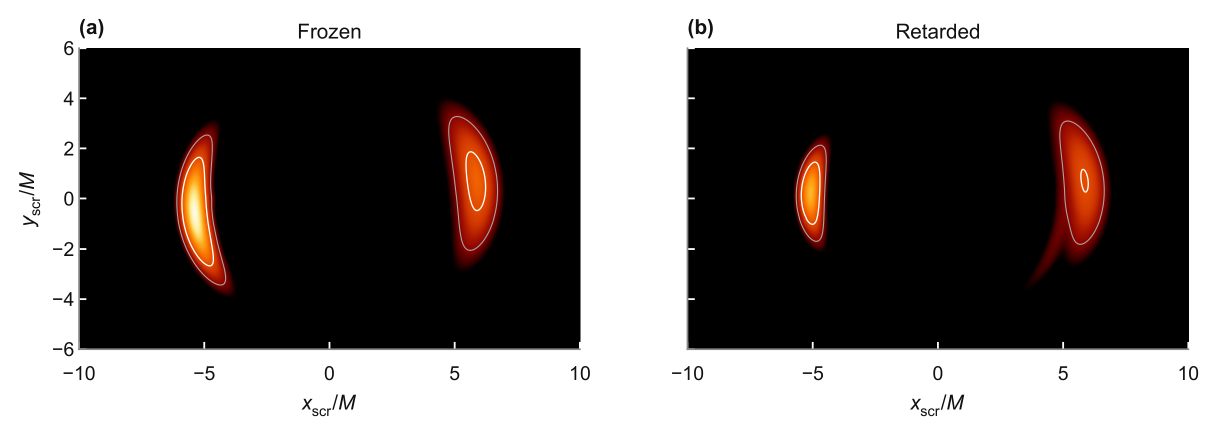}
\caption{High resolution expansion at \(\psi = 2.5\, rad.\) (a, b) Frozen and retarded images at \(N = 2048,\) normalized to the same frozen peak over the logarithmic range \(10^{-3} \leq I/I_{fr,peak} \leq 1.\) Gray and white contours mark \(0.01\) and \(0.1\) of that peak, respectively. Table 2 reports the regional intensity fractions and covariance budgets. Figure 10 shows the corresponding endpoint weights and budgets at \(N = 2048.\)}
\label{fig:paper64-9}
\end{figure}

\begin{figure}[htbp]
\centering
\includegraphics[width=\textwidth,height=0.76\textheight,keepaspectratio]{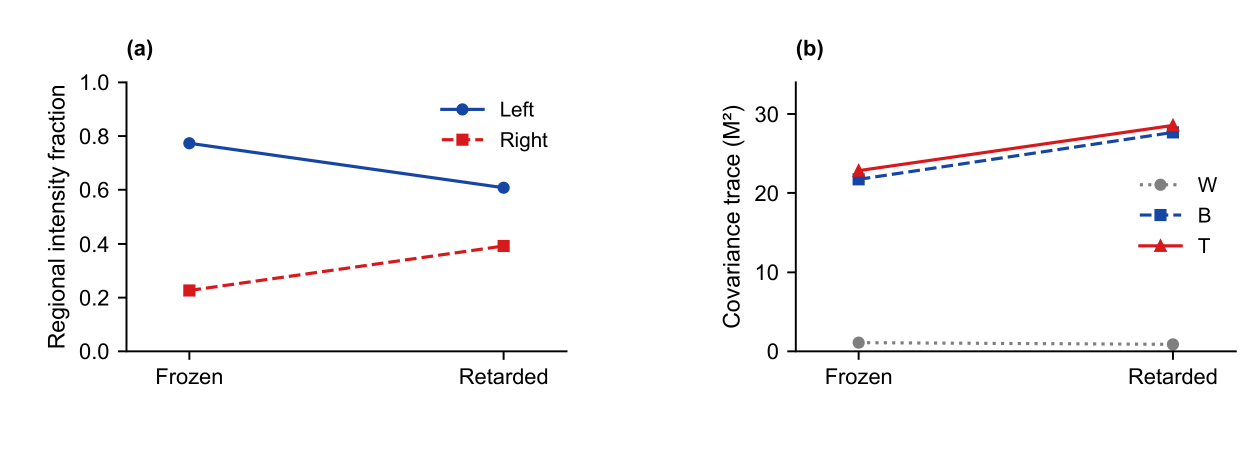}
\caption{Expansion endpoints at \(\psi = 2.5\, rad\) and \(N = 2048.\) (a) Left and right regional intensity fractions for the partition \(x = 0.\) (b) Covariance budgets \(W,\) \(B\) and \(T\) from the native arrays, using the same images as Fig.~9 and Table 2.}
\label{fig:paper64-10}
\end{figure}

The response ratio and \(\Delta B\) vary little across the tested resolutions (Appendix B.5).

For \(N=512\), the regional weighted median propagation delays differ by \(14.766\,M\), corresponding to approximately \(1.43\,rad\) of source advance. This associates the two regions with different portions of the trajectory; Appendix B.5 records the absolute medians and delay convention.

\subsection{Effects of lens and source geometry on redistribution}

Regional response factors depend on the lens and source geometry. Figure 11 compares three frozen configurations at the same source position and local static frame velocity. Because the principal and zero coupling Bardeen calculations have identical instantaneous source states and photon metrics, their images agree to roundoff; the Schwarzschild result provides a frozen background reference. Inclination and source size effects are characterized separately by the independent frozen and retarded image pairs in Fig.~12.

\begin{figure}[htbp]
\centering
\includegraphics[width=\textwidth,height=0.76\textheight,keepaspectratio]{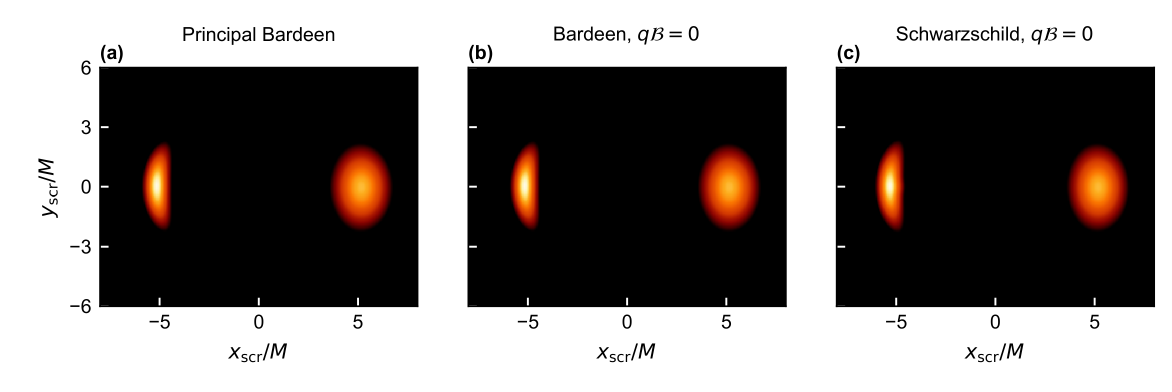}
\caption{Frozen background metric controls for \(\psi = 1.5\, rad\) at matched source position and local velocity. (a) Principal Bardeen; (b) Bardeen with \(q\mathcal{B} = 0;\) (c) Schwarzschild with \(q\mathcal{B} = 0.\) The native \(512 \times 512\) arrays are displayed after applying the same \(0.2\, M\) convolution. Each panel is normalized to its own peak over the range \(10^{-3} \leq I/I_{peak} \leq 1\).}
\label{fig:paper64-11}
\end{figure}

Lowering the inclination brings the frozen and retarded image morphologies closer (Fig.~12). With its total comoving emissivity matched to the principal source, the smaller hotspot retains a strong response at high inclination. The corresponding correlation, intensity and relative \(L_{2}\) measures are collected in Appendix B.5.

Regional redistribution remains visible in these retarded calculations, with its amplitude and small scale morphology varying across the tested inclination and source size configurations.

\begin{figure}[!t]
\centering
\includegraphics[width=\textwidth,height=0.76\textheight,keepaspectratio]{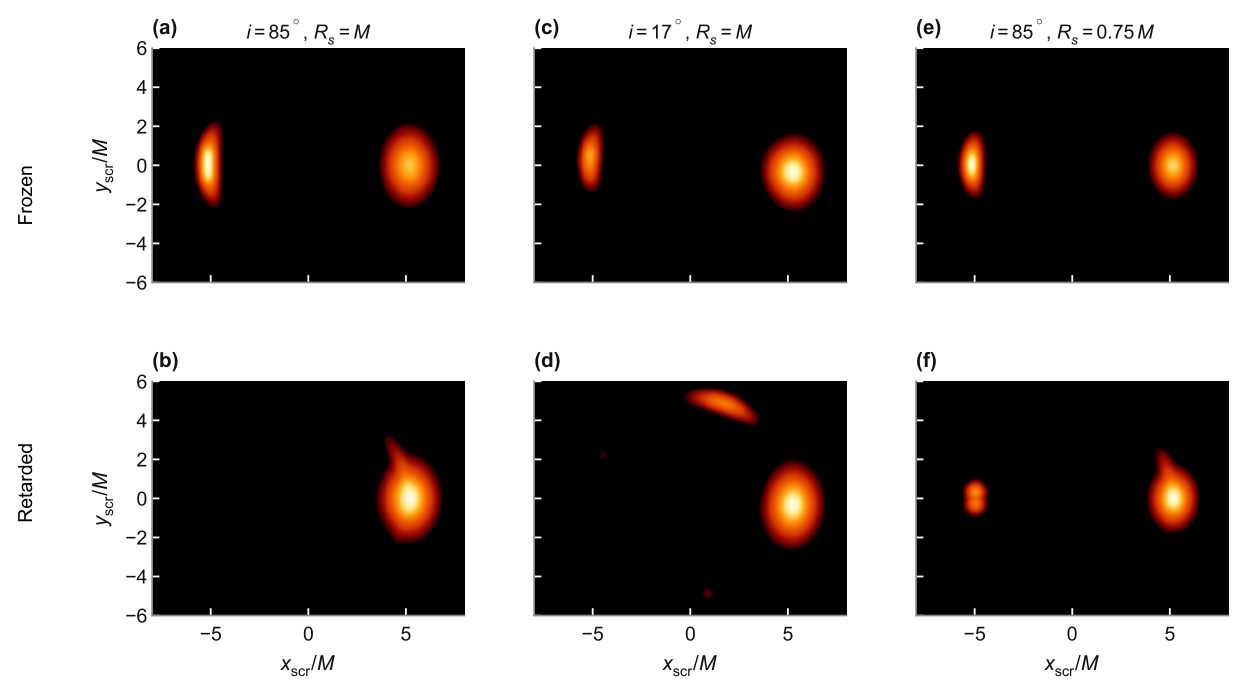}
\caption{Inclination and source size comparisons for the \(\psi = 1.5\, rad\) event. Columns show the principal configuration at \(N = 256\) (a, b), the lower inclination configuration at \(N = 256\) (c, d), and the smaller source configuration at \(N = 512\) (e, f). Frozen and retarded images occupy the two rows; the first column shows the uniform grid control. Each panel is normalized to its own peak, with a common field of view and range \(10^{-3} \leq I/I_{peak} \leq 1.\) Intensity changes are calculated from the underlying arrays.}
\label{fig:paper64-12}
\end{figure}

\makeatletter
\let\paperSavedSecFB\@fb@secFB
\let\@fb@secFB\relax
\section{Discussion}
\let\@fb@secFB\paperSavedSecFB
\makeatother

With the emitting plasma represented by a prescribed source profile, the second moment size characterizes the apparent spatial extent of the lensed intensity distribution and depends jointly on the widths of the bright regions, their relative weights and the separations of their centroids. In both events studied here, the within region contribution changes only slightly or varies in the opposite direction to the total covariance trace. Centroid and size changes measure different moments of the same regional redistribution, but their dependence on the regional weights differs because Eq. (2.19) is linear in \(p' - p\), whereas the between region covariance depends on \(p(1 - p)\). Across the two events, comparable dimming and decreasing regional centroid separation accompany opposite size responses because the regional intensity weights move in opposite directions relative to balance.

For multiple regions, Eqs. (2.12) and (2.15) replace the scalar balance curve by the weighted covariance of the regional centroids. Regional integrated intensities, centroids and covariances allow changes in regional weights to be separated from deformation within the regions.

A spatially resolved delay diagnostic preserves the association between emission time and screen position that is lost under screen averaging. Measured regional response factors recover the weight changes, and transfer histories connect these changes directly to the hotspot's position and velocity along its worldline. We use the response factors measured from completed frozen and retarded image pairs to interpret those pairs. Predictive applications require response factors obtained independently of the target retarded image, either from transfer histories or from a calibrated response model. These spatially resolved quantities complement time dependent transfer studies of light curves and variability \cite{Bronzwaer2018RAPTOR,Dexter2009,Pelle2022Skylight}, as well as order resolved studies relating lensing delays to distinct image families \cite{RojasPaternina2026,Kocherlakota2024}. Hotspot images and centroid tracks have been used to model source motion and constrain orbital geometry \cite{BroderickLoeb2006,Vincent2011,Baubock2020,Matsumoto2020Pattern}. Regional intensity redistribution can therefore contribute to apparent centroid and size variations when hotspot images are used to infer source motion.

The numerical examples examine two phases of the prescribed trajectory. Complete regional transfer histories support the analysis of the contracting event. Image moments and a regional delay summary characterize the expanding event. A gray, optically thin Gaussian represents the source. The geometry comparisons cover the selected background, inclination and source size configurations.

\section{Conclusion}

For a fixed observer time, differences in propagation delay cause separated image regions to sample different stages of an evolving hotspot's history. Regional response factors quantify the resulting changes in intensity weights. The total covariance separates these changes from variations in regional width and centroid separation. For two regions at fixed centroid separation, moving the weights away from balance reduces \(p(1 - p)\left| \mathbf{d} \right|^{2}\), whereas moving them toward balance increases it. Changes in centroid separation and the within region term \(\Delta W\) determine whether the total scale response follows this weight driven tendency.

Comparable dimming accompanies opposite size responses in the two events. At \(\psi = 1.5\), the image regions sample different intervals of the source history, and the initially near balanced intensity weights become strongly concentrated on one side of the image. At \(\psi = 2.5\), the initially weaker image region gains relative weight as the regional centroids move closer. The image expands because the contribution from regional intensity weights exceeds the opposing geometric contribution.

A lensed hotspot with fixed comoving source size can therefore appear smaller or larger when propagation delays change the relative intensity contributions of its separated image regions.

\appendix
\small

\section{Model details}

The prescribed vector potential \(A_{\phi}\) is evaluated on the Bardeen background. Photon and source separations are evaluated on the equal coordinate time slice of the worldline query. At the source center, the local static frame gives \(\xi^{\widehat{r}}=\sqrt{g_{rr}}\,\Delta r\), \(\xi^{\widehat{\theta}}=\sqrt{g_{\theta\theta}}\,\Delta\theta\), and \(\xi^{\widehat{\phi}}=\sqrt{g_{\phi\phi}}\,\Delta\phi\), with the shortest wrapped \(\Delta\phi\). The velocity \(v^{\widehat{\imath}}\) enters the comoving correction in Eq. (2.4). At emission, its components are converted from the local static frame to coordinate components, and \(u_\mu u^\mu=-1\) fixes the four velocity normalization. Comoving emissivity integration uses the Euclidean tangent space measure; the truncated Gaussian radial integral is analytic.

The prescribed field has the following Maxwell residual, whose nonzero value represents the current required to maintain the test field on the fixed Bardeen background:

\setcounter{equation}{0}
\begin{equation}
\nabla_{\mu}F^{\mu\phi} = \frac{\mathcal{B}}{r^{2}}\left\lbrack f(r) + rf'(r) - 1 \right\rbrack = - \frac{6\mathcal{B}g_{m}^{2}}{(r^{2} + g_{m}^{2})^{5/2}}.
\end{equation}

With \(E\equiv-p_t\), the equatorial conditions \(\theta=\pi/2\) and \(p_\theta=0\) reduce the mass shell relation to

\setcounter{equation}{1}
\begin{equation}
E^{2} = f^{2}p_{r}^{2} + f\left\lbrack 1 + \frac{(L - q\mathcal{B}r^{2}/2)^{2}}{r^{2}} \right\rbrack.
\end{equation}

The production initial conditions are \(q=0.18499344038963297\), \(\mathcal B=0.2\), \(L=3.335\), and \(r_\star=4.32804807199348\,M\). They give \(E=0.898772205296495\) and \(p_{r,0}=+0.106899999999850\) at \(r_0=6.5\,M\). Entry is the first inward crossing of \(\sqrt{(r-r_\star)^2+p_r^2}=0.05\) in code normalized variables. There the azimuth is reset to zero while coordinate time is retained; the observer azimuth is also zero. The recorded \(\psi=1.5\) event has \(t/M=202.40878423962528\) and \(r/M=4.3501180094429719\).

Equation (2.6) uses the standard \(\mathcal{G}_{\infty}^{3}\) transformation of monochromatic specific intensity. With frequency independent \(j_{\nu}\) and no absorption, the images represent formal gray specific intensity maps in code units. A frequency specific or bolometric interpretation requires a specified emission spectrum and the corresponding transfer coefficients.

\section{Numerical sensitivity and transfer comparison}

\subsection{Ray and source integration}

A ray step is accepted only when all paired fifth and sixth order component estimates satisfy \(|u_5-u_6|/(|u_6|+10^{-15})\leq10^{-14}\). The nominal radial step is \(h(r)=5\times10^{-4}(r/r_{eh})^{1.8}\), with \(r_{eh}=1.78596778\,M\) for \(g_m=0.5\,M\). Rejected steps are quartered; accepted steps return to this prescription. Null Hamiltonian rejection uses \(10^{-8}\), with no further step size bound. The source worldline uses proper time step \(0.0125\,M\) and stores \((t, r, \theta, \phi, p_t, p_r, p_\theta, p_\phi)\). Binary search brackets each coordinate time query; all eight components are linearly interpolated, wrapping \(\phi\) and setting the time component to the query value. Conserved \(p_t\) fixes the local static frame velocity. If needed, one factor rescales its three components to restore the mass shell before reconstructing canonical momenta. Queries remain within the stored interval.

Six successive \(17\times17\) screen searches from the projected source center select the valid direct primary ray by minimum closest approach distance. When the distance minima are indistinguishable, a propagation delay term of relative weight \(10^{-8}\) selects the shortest unwound ray. The anchor must pass within \(0.02\,M\) of the source center and have positive delay.

\subsection{Composite grid construction and region masks}

At \(\psi=1.5\), a crop of the effective \(N=4095\) grid contains \(2731\times1639\) pixels at \(\Delta x=\Delta y=0.00732780\,M\). In the fixed refinement window \(-5.30532\leq x/M\leq-3.38544\), \(-3.50269\leq y/M\leq3.70787\), \(263\times985\) coarse pixels are replaced by a \(525\times1969\) calculation at \(0.00366390\,M\) spacing. Both states use nominal pixel areas. The resulting \(0.2408\%\) area deficit relative to the replaced coarse grid window changes the two reported percentages by less than 0.00005 percentage points when closed in post processing.

Over the same domain, coarse and fine sampling give contractions of 43.1\% and 44.5\% and the fractions of the trace change due to regional weight changes are 98.5\% and 98.6\%, respectively. The contraction amplitude is more resolution sensitive than this fraction.

The fixed refinement window and its complement define the unsmoothed left and right masks in both images and native transfer increments. At \(\psi=1.5\), the test includes \(0.25\,M\) rectangular shifts and expansion, hard and soft two seed Voronoi masks, and an eight neighbor watershed. The soft weight is \(m_L=\{1+\exp[(d_L-d_R)/(0.50\,M)]\}^{-1}\), where \(d_L, d_R\) are screen plane seed distances. Seven masks retain separated regions and give \(\Delta T_{w\mid fr}=-18.192\) to \(-18.197\,M^2\), including the within region term; their frozen regional reference is coarser than the principal composite endpoint.

Soft mask weights are nonnegative and sum to unity per pixel. Weighted full domain sums define regional moments; matching masks across states preserve Eq. (2.15). At \(\psi=2.5\), the principal boundary is \(x=0\), scanned over \(-2\,M\leq x_0\leq2\,M\).

\subsection{Morphology statistics and control matching}

For pairwise statistics, we use the Gaussian convolved field \({\widetilde{I}}_{a}\) normalized by its screen integral:

\setcounter{equation}{0}
\begin{equation}
{\widehat{I}}_{a,ij} = \frac{{\widetilde{I}}_{a,ij}}{\sum_{kl}^{}{\widetilde{I}}_{a,kl}\Delta A_{kl}},\quad\quad a \in \{1,2\}.
\end{equation}

For any field \(X\) on a comparison domain \(\Omega,\) define the area weighted mean and norm by

\setcounter{equation}{1}
\begin{equation}
\langle X\rangle_{\Omega} = \frac{\sum_{(i,j) \in \Omega}^{}X_{ij}\Delta A_{ij}}{\sum_{(i,j) \in \Omega}^{}\Delta A_{ij}},\quad\quad\left. \parallel X \right.\parallel_{2,\Omega} = \left\lbrack \sum_{(i,j) \in \Omega}^{}X_{ij}^{2}\Delta A_{ij} \right\rbrack^{1/2}.
\end{equation}

Pearson correlation and relative \(L_{2}\) difference are calculated as

\setcounter{equation}{2}
\begin{equation}
\rho_{\Omega} = \frac{\sum_{(i,j) \in \Omega}^{}({\widehat{I}}_{1,ij} - \langle{\widehat{I}}_{1}\rangle_{\Omega})({\widehat{I}}_{2,ij} - \langle{\widehat{I}}_{2}\rangle_{\Omega})\Delta A_{ij}}{\left. \parallel{\widehat{I}}_{1} - \langle{\widehat{I}}_{1}\rangle_{\Omega} \right.\parallel_{2,\Omega}\left. \parallel{\widehat{I}}_{2} - \langle{\widehat{I}}_{2}\rangle_{\Omega} \right.\parallel_{2,\Omega}},\quad\quad\epsilon_{L_{2}} = \frac{\left. \parallel{\widehat{I}}_{cmp} - {\widehat{I}}_{ref} \right.\parallel_{2,\Omega}}{\left. \parallel{\widehat{I}}_{ref} \right.\parallel_{2,\Omega}}.
\end{equation}

The magnetic coupling and metric controls hold the selected source position and three local static frame velocity components fixed. Canonical momenta are reconstructed with each control metric and potential. Matched frozen state image moments use unsmoothed native arrays; displayed fields are convolved and peak normalized. Morphology statistics follow Eqs. (B.1)--(B.3) on the active union in Eq. (3.1).

All differences, ratios, percentages and decompositions are calculated from full precision values and rounded only for reporting. Displayed components and totals may therefore differ by one unit in the last quoted digit.

\subsection{Transfer and resolution validation}

For the \(\psi=1.5\) event, the stored worldline spans \(200.014\,M\) before and \(50.001\,M\) after the labeled state, covering the retarded sampling interval (Fig.~13). The normalized particle mass shell and photon null Hamiltonian residuals stay below \(4.44\times10^{-16}\) and \(3.47\times10^{-11}\). Halving the source history step changes the retarded images by relative \(L_2=4.37\times10^{-9}\) and \(1.75\times10^{-9}\) at \(\psi=1.5\) and \(2.5\); changing quadrature gives \(10^{-5}\) to \(10^{-4}\). At \(\psi=2.5\), raising \(N\) from 1024 to 2048 moves the retarded centroid by \(2.55\times10^{-4}\,M\) and changes the second moment size by \(5.73\times10^{-6}\) relatively.

\begin{figure}[htbp]
\centering
\includegraphics[width=0.85\textwidth,keepaspectratio]{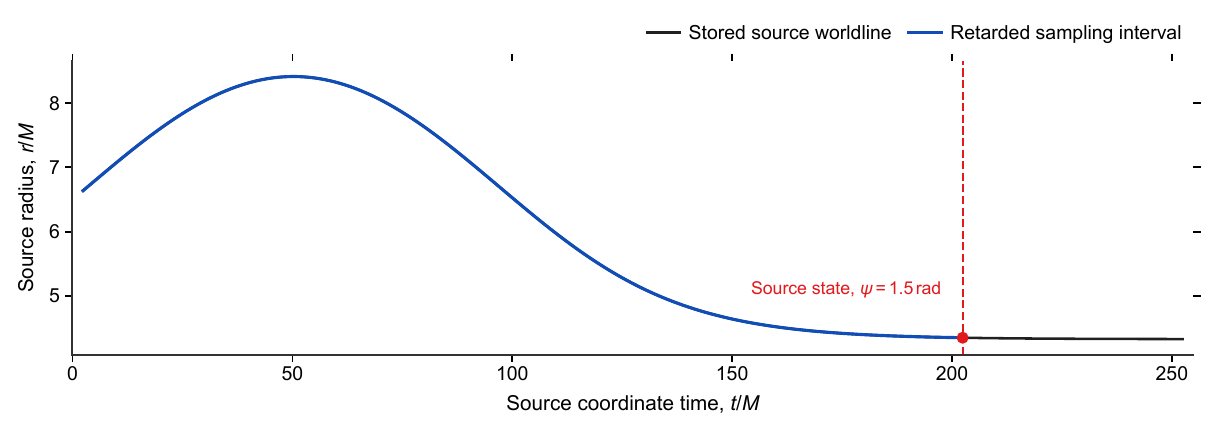}
\caption{Stored source worldline for the \(\psi=1.5\) event. The black curve shows the source radius over the stored trajectory, the blue interval extends from the earliest stored state to the labeled state and covers the retarded queries, and the red line and point mark the labeled state at \(t/M=202.409\). Coordinate time is retained from the production integration.}
\label{fig:paper64-13}
\end{figure}

In the flat spacetime benchmark, the source starts at \(\boldsymbol z_s(0)=(6M, 0, 0)\) with constant Cartesian velocity \((0, 0.15, 0)\), so the retarded calculation uses the inertial trajectory \(\boldsymbol z_s(t)=(6M, 0.15t, 0)\); the frozen calculation uses the state at \(t=0\). The source radius is \(R_s=M\). The camera is at \((20M, 0, 0)\), the screen covers \([-3M, 3M]^2\), and \(t_{obs}=14M\). A screen ray is parameterized by \(\boldsymbol z_\gamma(\ell)=(20M-\ell, x_{scr}, y_{scr})\) with \(t_{em}=14M-\ell\). The associated future directed photon momentum is \(k^\mu=-dx^\mu/d\ell\), so \(k^t=1\). The independent reference evaluates the same truncated comoving emissivity as \(I_D(x_{scr}, y_{scr})=\int_{8M}^{20M}\mathcal G_\infty^3j_\nu|k_\mu u^\mu|\,d\ell\). Taking \(I_D\) as the reference and the ray traced field \(I_R\) as the comparison, we measure errors by

\setcounter{equation}{3}
\begin{equation}
e_{F} = \frac{\left| F_{R} - F_{D} \right|}{F_{D}},\quad\quad e_{2,raw} = \frac{\left. \parallel I_{R} - I_{D} \right.\parallel_{2}}{\left. \parallel I_{D} \right.\parallel_{2}},\quad\quad e_{1,shape} = \sum_{ij}^{}\left| \frac{I_{R,ij}}{F_{R}} - \frac{I_{D,ij}}{F_{D}} \right|\Delta A.
\end{equation}

Photon and source step tests use \(96\times96\) grids; the intensity pair, residual, and profile difference use \(512\times512\) grids. For the latter, \(e_F\approx5.44505\times10^{-4}\), \(e_{2,raw}\approx5.43467\times10^{-4}\), and \(e_{1,shape}\approx4.47204\times10^{-6}\). Unit flux normalization separates the shape error from the dominant amplitude difference. Image convergence is approximately first order at the two smallest retained photon steps.

\begin{figure}[htbp]
\centering
\includegraphics[width=0.85\textwidth,keepaspectratio]{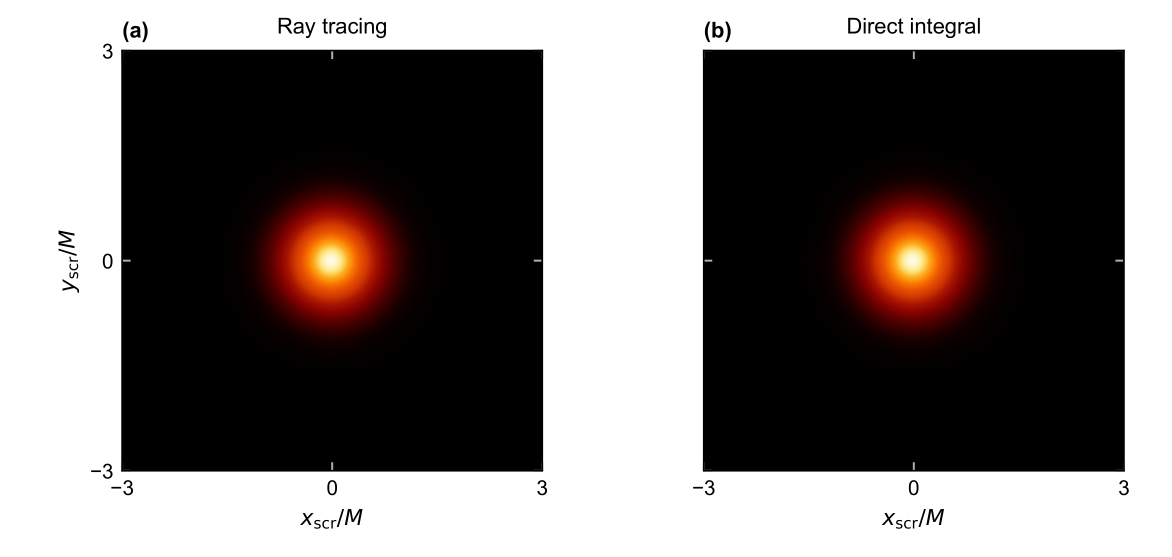}
\caption{Native \(512\times512\) retarded flat spacetime benchmark. (a) Ray traced field \(I_R\); (b) independent direct integral field \(I_D\). Both panels are normalized to \(\max I_D\), use the same linear range \(0\leq I/\max I_D\leq1\), and are shown without smoothing. The steps are \(h_\gamma=5\times10^{-4}M\) and \(h_s=10^{-2}M\).}
\label{fig:paper64-14}
\end{figure}

\begin{figure}[htbp]
\centering
\includegraphics[width=0.78\textwidth,keepaspectratio]{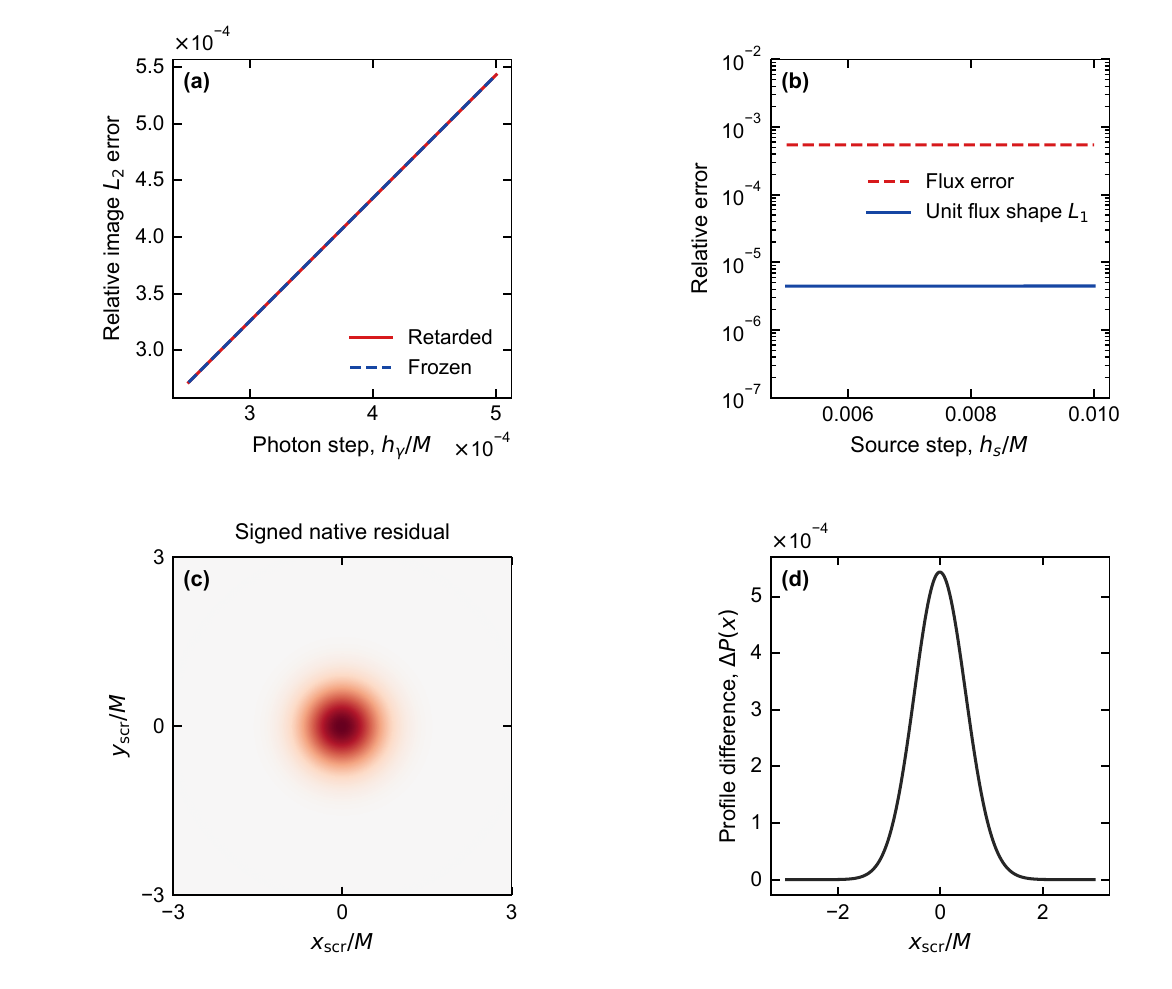}
\caption{Flat spacetime validation against the independent direct integral. (a) Relative image \(L_2\) error versus photon step for retarded and frozen transfer; the retarded calculation uses \(h_s=10^{-2}M\). (b) Flux error \(e_F\) and unit flux shape error \(e_{1,shape}\) versus source step at fixed \(h_\gamma=5\times10^{-4}M\). Panels (a,b) use \(96\times96\) grids. (c) Native \(512\times512\) residual \(R=(I_R-I_D)/\max I_D\), shown without smoothing on the symmetric range \(|R|\leq5.42\times10^{-4}\). (d) Profile difference \(\Delta P=P_R-P_D\), where \(P_X(x)=\int I_X(x,y)\, dy/\max_x\int I_D(x,y)\,dy\).}
\label{fig:paper64-15}
\end{figure}

Under the direct ray anchor, attenuation and contraction persist at \(\psi=1.5\) for \(-0.5\leq\Delta\psi\leq0.5\,rad\). At \(\psi=2.5\), an independent retarded anchor changes the image by relative \(L_2=0.001716\) and the centroid by \(1.99\times10^{-3}\,M\), relative to the production anchor.

\subsection{Additional endpoint diagnostics}

For \(\psi=(1.5, 2.5)\), the major covariance eigenvalue changes are \((-18.506, +5.972)\,M^2\), and the minor changes are \((+0.057, -0.244)\,M^2\); both major axis rotations are below \(4.3^\circ\). At \(\psi=1.5\), the regional median delay separation corresponds to \(3.32\,rad\). Centroid shifts from regional weights alone and from the full calculation are \(3.853\) and \(3.854\,M\); their active region image difference is relative \(L_2=0.103\) with the reconstruction as reference. Within region rms sizes change by \((-2.56, +5.64)\%\).

At \(\psi=2.5\), \(N=(512, 1024, 2048)\) gives response ratios \((2.171, 2.202, 2.203)\) and \(\Delta B=(5.901, 5.936, 5.945)\,M^2\). With \(\tau_{prop}=-t_\gamma\), the \(N=512\) regional medians are \((1032.454, 1017.689)\,M\), differing by \(14.766\,M\) or \(1.43\,rad\) of source advance.

At \(N=256\), \(i:85^\circ\to17^\circ\) gives correlation \(0.466\to0.928\), relative \(L_2:0.925\to0.322\), and intensity response \(-61.8\%\to-18.0\%\). For the smaller \(R_s=0.75\,M\) hotspot, the total comoving emissivity is matched to the principal source. Its intensity changes are \(-62.0\%\) at \(N=256\) and \(-57.2\%\) at \(N=512\).

\FloatBarrier
\normalsize
\section*{Acknowledgments}

This work is supported in part by the Basic Scientific Research Operating Expenses Program of China West Normal University under Grant No. 2026kx007.



\bibliographystyle{JHEP}
\bibliography{refs}
\end{document}